\documentclass[aps,pra,reprint,superscriptaddress,showkeys,nofootinbib,floatfix]{revtex4-1}

\usepackage[T1]{fontenc}			
\usepackage[utf8]{inputenc}			
\usepackage[english]{babel}			
\usepackage{microtype}				

\usepackage{color} 						
\definecolor{red}{rgb}{1,0,0}					
\definecolor{blue}{rgb}{0,0,1}					
\definecolor{black}{rgb}{0,0,0}				

\usepackage{soul} 
\definecolor{hlyellow}{rgb}{0.95,0.95,0}
\definecolor{hlgreen}{rgb}{0,0.95,0}
\sethlcolor{hlyellow}
\soulregister \ref 7
\soulregister \cite 7
\soulregister \citet 7
\soulregister \footnote 8
\soulregister {\ } 7
\soulregister \figref 7
\soulregister \tabref 7
\soulregister \secref 7
\soulregister \refref 7

\usepackage{amsmath}	
\usepackage{amssymb,amsfonts,mathrsfs} 
\usepackage{array} 

\usepackage{graphicx}			

\usepackage{hyperref}  
\definecolor{dullmagenta}{rgb}{0.4,0,0.4}   
\definecolor{darkblue}{rgb}{0,0,0.4}
\definecolor{medblue}{rgb}{0,0,0.6}
\definecolor{lightblue}{rgb}{0,0,0.8}
\hypersetup{
	colorlinks=true, 		
	breaklinks=true,		
	linkcolor=lightblue,
	citecolor=lightblue,
	urlcolor=lightblue,
	filecolor=dullmagenta
}
\usepackage[all]{hypcap} 	

\usepackage{physmaths}
\usepackage{physunits}

\newcommand{\figscale}{1} 
\newcommand{\figref}[1]{Fig.\ \ref{#1}} 
\newcommand{\tabref}[1]{Tab.\ \ref{#1}} 
\newcommand{\secref}[1]{Sec.\ \ref{#1}} 
\newcommand{\figletter}[1]{\textbf{(\mbox{#1})}}
\newcommand{\refref}[1]{Ref.\ \cite{#1}} 

\newcommand{\DBz}{\Delta B_z}
\newcommand{\tC}{t_\text{C}}

\begin{document}

\newcommand{\mytitle}
{High-fidelity single-shot readout for a spin qubit via an enhanced latching mechanism}
\title{\mytitle}

\author{Patrick \surname{Harvey-Collard}}
\email[Correspondance to: ]{P.Collard@USherbrooke.ca}
\affiliation{Département de physique et Institut quantique, Université de Sherbrooke, Sherbrooke, QC, J1K 2R1, Canada}
\affiliation{Sandia National Laboratories, Albuquerque, NM, 87185, United States}

\author{Benjamin \surname{D'Anjou}}
\affiliation{Department of Physics, McGill University, Montréal, QC, H3A 2T8, Canada}

\author{Martin \surname{Rudolph}}
\affiliation{Sandia National Laboratories, Albuquerque, NM, 87185, United States}

\author{N. Tobias \surname{Jacobson}}
\affiliation{Center for Computing Research, Sandia National Laboratories, Albuquerque, NM, 87185, United States}

\author{Jason \surname{Dominguez}}
\affiliation{Sandia National Laboratories, Albuquerque, NM, 87185, United States}
\author{Gregory A. \surname{Ten~Eyck}}
\affiliation{Sandia National Laboratories, Albuquerque, NM, 87185, United States}
\author{Joel R. \surname{Wendt}}
\affiliation{Sandia National Laboratories, Albuquerque, NM, 87185, United States}
\author{Tammy \surname{Pluym}}
\affiliation{Sandia National Laboratories, Albuquerque, NM, 87185, United States}

\author{Michael P. \surname{Lilly}}
\affiliation{Center for Integrated Nanotechnologies, Sandia National Laboratories, Albuquerque, NM, 87185, United States}

\author{William A. \surname{Coish}}
\affiliation{Department of Physics, McGill University, Montréal, QC, H3A 2T8, Canada}
\affiliation{Quantum Information Science Program, Canadian Institute for Advanced Research, Toronto, ON, M5G 1Z8, Canada}
\affiliation{Center for Quantum Devices, Niels Bohr Institute, University of Copenhagen, 2100 Copenhagen, Denmark}

\author{Michel \surname{Pioro-Ladrière}}
\affiliation{Département de physique et Institut quantique, Université de Sherbrooke, Sherbrooke, QC, J1K 2R1, Canada}
\affiliation{Quantum Information Science Program, Canadian Institute for Advanced Research, Toronto, ON, M5G 1Z8, Canada}

\author{Malcolm S. \surname{Carroll}}
\email[Correspondance to: ]{mscarro@sandia.gov}
\affiliation{Sandia National Laboratories, Albuquerque, NM, 87185, United States}

\date{December 8$^\text{th}$, 2017}

\begin{abstract}
The readout of semiconductor spin qubits based on spin blockade is fast but suffers from a small charge signal. Previous work suggested large benefits from additional charge mapping processes, however uncertainties remain about the underlying mechanisms and achievable fidelity. In this work, we study the single-shot fidelity and limiting mechanisms for two variations of an enhanced latching readout. We achieve average single-shot readout fidelities $>99.3\%$ and $>99.86\%$ for the conventional and enhanced readout respectively, the latter being the highest to date for spin blockade. The signal amplitude is enhanced to a full one-electron signal while preserving the readout speed. Furthermore, layout constraints are relaxed because the charge sensor signal is no longer dependent on being aligned with the conventional $(2,0)-(1,1)$ charge dipole. Silicon donor-quantum-dot qubits are used for this study, for which the dipole insensitivity substantially relaxes donor placement requirements. One of the readout variations also benefits from a parametric lifetime enhancement by replacing the spin-relaxation process with a charge-metastable one. This provides opportunities to further increase the fidelity. The relaxation mechanisms in the different regimes are investigated. This work demonstrates a readout that is fast, has one-electron signal and results in higher fidelity. It further predicts that going beyond $99.9\%$ fidelity in a few microseconds of measurement time is within reach. 

\end{abstract}

\maketitle


\section{Introduction}

There is a rapidly growing commercial interest in quantum computing for applications such as optimization and quantum chemistry.  A number of companies are now attempting to build small quantum bit (qubit) \cite{loss1998,kane2000} platforms for conceptual testing.  Quantum dot spin qubits are of interest because of their promising coherence properties, the solid-state all-electrical control that can be achieved and the potential to be built on the semiconductor fabrication platform already used for high performance computing.  
Qubit control fidelities have been studied extensively and reached relatively low error probabilities \cite{veldhorst2014a,veldhorst2015a,muhonen2015,kawakami2016a,takeda2016a,nichol2017}. However, state preparation and readout errors have yet to reach similarly low error levels \cite{barthel2009,shulman2012a,higginbotham2014b,bertrand2015,nichol2017}. Even though fault tolerance thresholds lie at the $1\pc$ level for one error correction round, individual components need to be much better (approximately $0.1\pc$ error probability or better).

Spin qubit states can be measured using a spin-to-charge conversion mechanism that maps spin states to charge states using Pauli spin blockade, followed by readout with a charge sensor (CS) \cite{johnson2005}. The minimum achievable error in this readout depends fundamentally on two time scales: the time needed to accurately distinguish between two readout states and the lifetimes of those states. For instance, to achieve a $10^{-3}$ error probability, the measurement time should be roughly $10^{3}$ times shorter than the signal lifetime. Readout speed is also a concern for the long-term viability of semiconductor spin qubits. Readouts based on energy-selective tunnelling events \cite{elzerman2004,tracy2016,watson2017,broome2017} have been shown to achieve $< 1 \pc$ error probabilities. However, increasing the signal lifetime and fidelity requires a reduced tunnel rate, which also makes the readout slower. In contrast, the conventional spin-blockade readout generates signal right away but suffers from a smaller charge dipole signal instead of a full one-electron signal \cite{johnson2005}. This adversely affects the readout speed and layout constraints.

Previous work has established that it is possible to improve the signal amplitude using mappings to metastable charge states \cite{petersson2010,studenikin2012a,mason2015}. However, some key questions remain. First, a single-shot readout using this enhancement process has not been demonstrated. It is not clear that the charge mapping process can be achieved with low-enough error rate to achieve high-fidelity single-shot readout \cite{studenikin2012a}. Second, it has not been demonstrated whether the signal lifetime could also be enhanced via this charge mapping process. It was reported that the expected signal enhancement was not observed in one case \cite{mason2015}, raising doubts as to whether this process can be achieved with high fidelity. Third, the mechanisms that limit the lifetimes of the metastable states are still not understood, and a comprehensive comparison of various alternative mapping schemes is lacking.

In this work, we study two variations of an enhanced latching readout mechanism, quantify the fidelity enhancement for single-shot readout and clarify some important error mechanisms. The work is performed using a silicon quantum dot (QD) coupled to a single donor (D) atom \cite{harvey-collard2017a}. The two-electron QD-D system can be thought of as a singlet-triplet (ST) qubit \cite{levy2002,petta2005,maune2012} in an effective double QD configuration \cite{taylor2007} where only one of the QDs is connected directly to a charge reservoir \cite{harvey-collard2017a}. This configuration produces latching (or hysteresis \cite{yang2014a}) of the QD-D charge state that can be harnessed to enhance the charge detection \cite{petersson2010,studenikin2012a,mason2015} in two ways. 

First, the spin states can be mapped to charge configurations that differ by one electron. Compared to the small dipole produced by the traditional readout, this creates a much higher charge signal that is very easily detected by the CS. We show that the CS then doesn't need to be aligned with this charge dipole, which enables detection in configurations where the traditional CS signal would vanish. This has profound implications in terms of design, particularly for QD-D systems and multi-qubit systems where conflicting layout constraints add-up \cite{watson2015}. 

Second, the latching behavior can extend the lifetime of the charge signal by orders of magnitude by changing the spin relaxation mechanism to a metastable charge relaxation one. Such an improvement can drastically improve detection and could take fast single-shot readout fidelity well into the (putative) fault tolerant threshold regime. The improvement could be particularly pronounced in materials like GaAs, where the spin blockade lifetime is $\sim 10 \us$ \cite{barthel2012}. Previous work has identified slow tunneling or co-tunneling processes as limiting the metastable charge lifetimes \cite{mason2015,yang2014a}. Our work identifies a parametric dependence expected for a mechanism based on the hybridization between the two effective QDs as the factor limiting the lifetime in the readout region, allowing us to identify an optimized readout regime. We identify two variations of the enhancement process. While both share the signal enhancement property, only one features the parametric lifetime improvement.

We develop a model of the readout mechanisms. This model is then used to analyze the single-shot experiments and demonstrate that the enhancement process can lead to an improvement in readout fidelity. We directly compare the benefits of the enhanced latching readout with those of the traditional spin-blockade readout by breaking down errors into sequential processes that add together. We leave out errors that could occur during the transit from separated electrons to the spin blockade region. These errors are studied in other work \cite{taylor2007,barthel2009,barthel2012,cerfontaine2014}, are common to all readouts and can be made sufficiently small. We account for mapping errors from the additional enhancement processes and from the final CS measurement. We use these techniques to demonstrate a readout fidelity $> 99.86\pc$ in $65\us$, the highest reported for spin blockade so far. Finally, it is worth noting that the results discussed in this work apply not only to donors but to general ST and all-exchange qubit systems where such a charge latching effect can be engineered. These include Si \cite{dehollain2014a,weber2014,veldhorst2015a}, Si/SiGe \cite{maune2012,wu2014a,eng2015} and GaAs/AlGaAs \cite{petta2005,foletti2009,gaudreau2012,medford2013b}.

\section{Results}

\subsection{Experimental system}

\begin{figure}
   \centering
   \includegraphics[scale=\figscale]{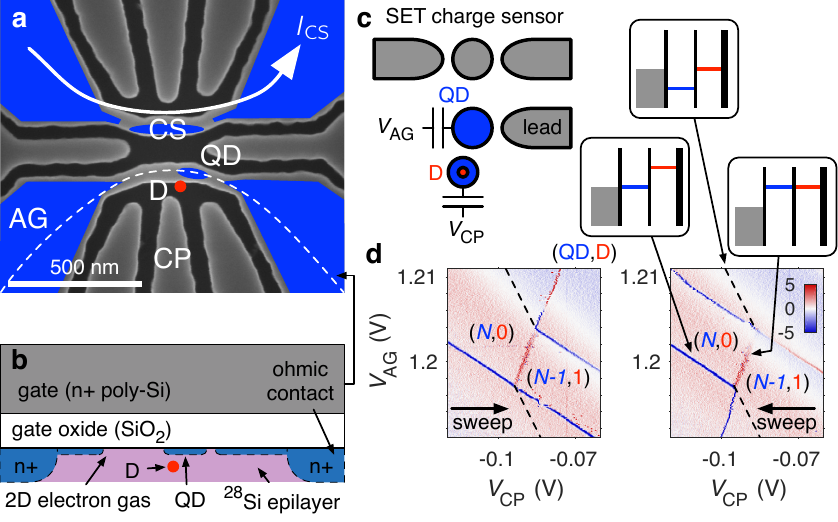} 
   \caption{\figletter{a} Scanning electron microscope image of the gate structure used to define the QD and CS. The blue overlay indicates the approximate shape of the electron gas. The QD is pushed to the right side and tunnel-coupled to a single reservoir. Phosphorus donors have been implanted in a self-aligned way at the location indicated by the red dot. Some donors are tunnel-coupled to the QD. Gates CP and AG are used to control the effective double-QD through voltages $V_\text{CP}$ and $V_\text{AG}$.     \figletter{b} MOS device gate stack along the dashed line of (a).     \figletter{c} The QD-D system effectively forms a double-QD-like system where D states have $0$ or $1$ electrons and have little or no coupling to a charge reservoir.    \figletter{d} Charge anti-crossing between a QD and a D state. The absence of reservoir for the D makes the charge states latch, which is made apparent by reversing the sweep direction of the scan. The dashed lines indicate where the equilibrium D\ $\leftrightarrow$\ lead transition should be. Colour scale: d$I_\text{CS}$/d$V_\text{CP}$ (a.~u.). }
   \label{fig:experiment}
\end{figure}

The experiments are performed in a silicon metal-oxide-semiconductor (MOS) QD-D device in a dilution refrigerator \cite{harvey-collard2017a}. A patterned poly-Si gate structure is used to confine electrons at the Si--\ce{SiO2} interface and is shown in \figref{fig:experiment}(a-b). 
The device is electrically biased to form a single electron transistor (SET) in the upper wire that is used as a CS, and a few-electron QD in the lower wire. The QD is asymmetrically biased such that it is coupled to a single reservoir. The resulting system is shown schematically in \figref{fig:experiment}(c). Phosphorus donors have been implanted in a self-aligned way at the location indicated by the red dot. See the ``Methods'' section for details. Some of the implanted donors are tunnel-coupled to the QD and together with it form an effective double-QD-like system where D states can accommodate a limited number of electrons (e.g.\ $0$, $1$). 

The donors in this work are weakly coupled to charge reservoirs. This inhibits relaxation to the charge ground state via direct tunnelling. As a result, donor electrons can instead go through the QD to exchange with the lead (see e.g.\ \refref{yang2014a} for a double QD version), which can also be relatively slow. This gives rise to a charge latching or hysteresis effect in charge stability diagrams, as shown in \figref{fig:experiment}(d). Here we denote QD-D charge states by $(N_\text{QD},N_\text{D})$, where $N_\text{QD}$ and $N_\text{D}$ are the number of electrons on the QD and D, respectively.

The system is tuned around a $(4,0)-(3,1)$ QD-D transition. All the experiments are realized in this four-electron charge configuration. The four-electron filling increases the ST readout window, presumably through valley shell filling. This donor-dot system behaves like an effective, spin-blockaded $(2,0)-(1,1)$ ST qubit and is described in detail in \refref{harvey-collard2017a}. Therefore, the two-electron notation is used throughout the text.

Two different donors are featured in this work. Donor 1 is featured in sections \ref{sec:readoutmech} through \ref{sec:relaxmech}. It has a smaller tunnel coupling ($\sim 0.5 \ueV$) that is well suited to study some of the relaxation physics detailed later. It also has all readout variations working simultaneously, allowing for a fair comparison of these variations. Donor 2 is featured in section \ref{sec:highFidelity}. It has a large tunnel coupling ($\gtrsim 20 \ueV$) and exhibits coherent behavior. It is used to demonstrate high single-shot readout fidelity.

\subsection{Readout mechanism}
\label{sec:readoutmech}

\begin{figure}
   \centering
   \includegraphics[scale=\figscale]{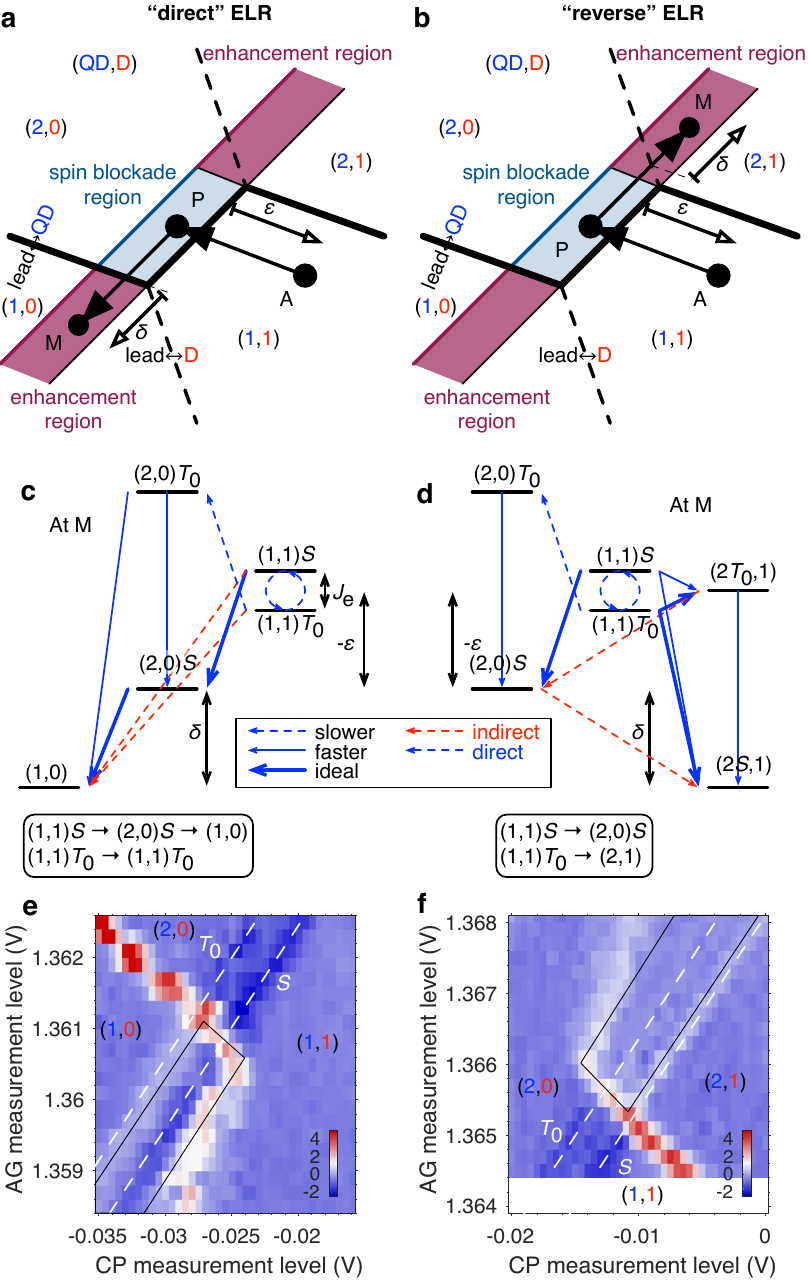} 
   \caption{\figletter{a-b} Two variations of the enhanced latching readout (ELR). In a typical experiment, one would control a ST qubit at point A, and pulse to point P in the Pauli spin blockade (PSB) region for PSB readout (PSBR). From there, an additional pulse to point M for measurement allows the selective mapping of one of the PSB states to a different charge occupation, therefore enhancing the amplitude of the sensed charge signal. Because the D\ $\leftrightarrow$\ lead tunnel rate is very small, the corresponding line in the charge stability diagram can be ignored. The thicker black lines represent relatively fast transitions. A nice way to understand the cause of the enhancement is that the excited state in the PSB region sees a shifted anti-crossing; therefore it does not cross the same QD\ $\leftrightarrow$\ lead transition lines as the ground state does when going from point P to point M.     \figletter{c-d} Energy levels and state ladder at point M. The $(1,1)S$ (or $(1,1){\uparrow\downarrow}$) is mapped to $(2,0)S$ through rapid (or slow) adiabatic passage from A to P. From P to M tunnelling processes with the lead perform the enhancement. The $\epsilon$ and $\delta$ parameters can be used to follow the energy detunings throughout the sequence. See main text section \ref{sec:conditionsELR} for details.     \figletter{e-f} A pulse sequence where the M point is swept is used to image the edges of the PSB and enhancement regions of donor 1. The measurement is initialized with a random state and averaged over many cycles. Colour scale: d$I_\text{CS}$/d$V_\text{CP}$ (a.~u.).}
   \label{fig:mechanism}
\end{figure}

We now show how the latching behavior of QD-D or QD-QD systems can be harnessed to produce a spin readout with very low error rate. To read out a ST qubit, one typically starts with a $(1,1)$ state, as shown in \figref{fig:mechanism}(a-b). Using fast voltage pulses on the device gates, the electron configuration is brought from point A in $(1,1)$ to point P in the Pauli spin blockade window \cite{johnson2005a,maune2012}. The $(1,1)S,T_0$ (or $(1,1){\uparrow\downarrow},\downarrow\uparrow$) states are mapped to either a $(2,0)S$ or an excited $(1,1)T_0$ by rapid (or slow) adiabatic passage \cite{taylor2007}. This process is known as Pauli spin blockade (PSB) spin-to-charge conversion. The CS measures the difference in charge configurations, although the net change of charge between the two readout states is zero. For this readout to produce a good signal, the CS needs to be somewhat aligned with the charge dipole. The signal lifetime is determined by the relaxation from the excited $(1,1)T_0$ to the $(2,0)S$ ground state and is a major factor limiting high readout fidelities. It is typically longer in Si than GaAs systems due to the absence of piezoelectric phonons at the relevant energies. 

In \figref{fig:mechanism}(a-b), we detail two variations of an enhanced latching readout (ELR), which we call the ``direct'' and ``reverse'' variations. The schematics depict a charge anti-crossing with the different charge regions identified. The black thick lines mark fast ground-state transitions between the QD\ $\leftrightarrow$\ lead and the QD\ $\leftrightarrow$\ D. The dashed D\ $\leftrightarrow$\ lead line does not play a role due to the charge latching and can be ignored in the following reasoning. The PSB region is contained between the $(2,0)S-(1,1)S$ inter-dot degeneracy line (black) and the $(2,0)T_0-(1,1)T_0$ one (blue). To take advantage of the charge enhancement, one can pulse the gate voltages from point P to one of the extensions of the PSB regions called ``enhancement regions'' at point M. Because such a pulse crosses one of the fast QD\ $\leftrightarrow$\ lead transition lines, the charge state is rapidly and conditionally mapped to the corresponding one depending on the state at P. This causes the total number of electrons to differ by one, which generates more signal than the PSB readout (PSBR) and is less geometry-dependent. 

\subsubsection{Direct ELR}

We define a direct ELR in which the $(2,0)S$ is mapped to a $(1,0)$ state by tunnelling with the lead, as shown in \figref{fig:mechanism}(c). The $(1,1)T_0$ is blocked from reaching the $(1,0)$ ground state through $(2,0)$ by the PSB, and cannot rapidly lose the D-side electron to the lead either due to the charge latching. The limiting factor for the signal lifetime is the same as for the PSBR. The signal amplitude corresponds to one additional electron on the donor instead of the $(2,0)-(1,1)$ dipole of the PSBR. We note that the term ``direct'' is coined because it is possible for the pulse trajectory to go directly from A to M without the detour by P. While optional for readout, this detour can be useful for qubit control or readout comparison (as in this work).

We can experimentally reveal the edges of the enhancement and PSB regions (i.e.\ readout windows) in \figref{fig:mechanism}(a) by preparing a random state at point A and varying the location of point M in an averaged measurement, as shown in \figref{fig:mechanism}(e). The random state is obtained by loading a $(2,0)S$, performing rapid adiabatic passage through the inter-dot transition, then waiting longer than the coherence time of the qubit. The voltage is pulsed to point P, then to point M. The location of point M is varied to image the charge regions. The time spent at M is the longest in the pulse sequence, and therefore the signal originates mostly from the charge state at M in this averaging mode. 

\subsubsection{Reverse ELR}

In the reverse variation, it is the $(1,1)T_0$ state that is mapped to a $(2,1)$ state by tunnelling with the lead, as shown in \figref{fig:mechanism}(d). The CS signal is again equivalent to a one-electron difference on the donor. In contrast with the direct ELR, this configuration has the significant advantage that the mechanism limiting the lifetime of the signal is transferred to a charge relaxation one that is no longer dependent on the traditionally-limiting PSB relaxation. This can be a significant advantage in any system, and particularly so in GaAs where the PSB lifetime poses significant challenges \cite{barthel2012}. This charge relaxation mechanism is discussed further in section \ref{sec:relaxmech}.

We experimentally reveal the edges of the enhancement region using a similar technique as in the direct case. The results are shown in \figref{fig:mechanism}(f). 

\subsubsection{Conditions for enhancement}
\label{sec:conditionsELR}

The state diagrams in \figref{fig:mechanism}(c-d) show important states involved at point M and the transitions between them. The solid lines are used for relatively fast processes, and the dashed lines for relatively slow ones. Blue lines link states involving direct transitions, while red lines represent indirect transitions suppressed by the weak D\ $\leftrightarrow$\ lead tunnel rate (typically Hz in this work). 

For accurate mapping, the QD $\leftrightarrow$ lead tunnel rate (typically MHz in this work) must be fast compared with the measurement time. In the direct ELR variation, a slow $(2,0)S \rightarrow (1,0)$ event can look like a fast $(1,1)T_0$ decay. In the reverse variation, a slow $(1,1)T_0 \rightarrow (2,1)$ rate can compete with the conventional PSB relaxation rate and introduce a branching process that limits the conversion efficiency. 

Indirect transitions can limit the metastable lifetime.  For the direct ELR variation, a small $(2,0)$ admixture in the $(1,1)$ metastable state can lead to a transition to the $(1,0)$ ground state via direct QD $\leftrightarrow$ lead tunnelling. For the reverse ELR variation, a small $(1,1)$ admixture in the $(2,0)$ metastable state can similarly lead to a transition to the $(2,1)$ ground state. Evidence for this mechanism is presented in \secref{sec:relaxmech}.

In the limit of a weak admixture mechanism, the metastable state in the direct ELR variation can still decay through the same triplet-singlet mechanism that limits the PSBR, following the path: $(1,1)T_0 \rightarrow (1,1)S \rightarrow (2,0)S \rightarrow (1,0)$.  In this case, the metastable lifetime for the direct ELR variation will be comparable to the lifetime of the triplet state for PSB decay, although this readout will still benefit from an improved signal contrast.  In contrast, in the reverse ELR variation, the metastable lifetime of the $(2,0)S$ state can be parametrically longer than the $(1,1)T_0$ lifetime for the PSBR, allowing for an improved lifetime in addition to an improved contrast.

Errors produced by the competition between the different intended and unintended transitions rates are discussed further later in the paper.

\subsection{Fidelity/error metric}
\label{sec:fidelitymetric}

We use the average readout fidelity $\bar F = 1-\bar e$ as a metric, where $\bar e = (e_S + e_T)/2$ is the average error probability for singlets and triplets. Since the scope of this work is to compare the benefits of the ELR with those of the traditional PSB readout, we account for errors that accumulate after the arrival at point P in the PSB window and neglect errors that could occur during the transit from $(1,1)$ to the PSB region. We account for the additional errors that can occur from P to M as a result of the added complexity and pulses required for the ELR. We call these mapping errors $\bar e_\text{map}$ (these do not apply to the PSB readout). After the pulse arrival at point M, the CS state discrimination process is started, which can also produce errors. We call these measurement errors $\bar e_\text{meas}$. For small errors, the total error $\bar e_\text{tot}$ from composed sequential processes can simply be added, $\bar e_\text{tot} = \bar e_\text{map} + \bar e_\text{meas}$ (see the Supplementary information section ``Error composition formula'').

\subsection{Direct comparison between readouts}
\label{sec:readoutcomp}

We use the donor 1 anti-crossing featured in \figref{fig:mechanism} to compare the characteristics of the readout variations. The data is acquired in a short period of lab time using the same nominal conditions to allow a fair comparison. Donor 1 had an inter-dot tunnel coupling large enough to allow adiabatic charge transfer ($\sim 0.5 \ueV$), but only with slow detuning ramps ($\sim 10 \us$). Nevertheless, it allows us to compare the PSBR and the two ELR variations under the same experimental conditions. In particular, we extract relaxation times, mapping errors, and signal enhancements. We do not report the measurement errors of the readout for donor 1. This is done in the subsequent section for donor 2.

\begin{figure}
   \centering
   \includegraphics{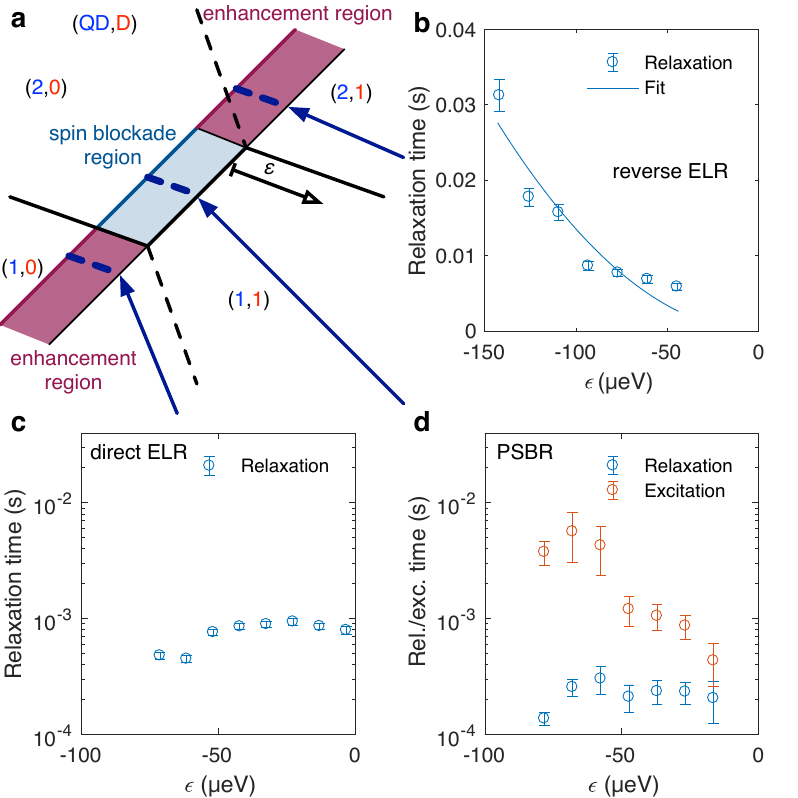} 
   \caption{\figletter{a-d} Relaxation/excitation time versus the detuning $\epsilon$ for the different readouts on donor 1. The data is taken along detuning cuts schematized in (a). Non-negligible excitation times are only observed in the PSBR case. For the ELR, excitation is strongly suppressed by the large $\delta > 100 \ueV$ energy cost (see \figref{fig:mechanism}). The fit in (b) corresponds to a simple relaxation model where the metastable relaxation rate $\varGamma_\text{MS} = \varGamma_{(2,1)\leftarrow(1,1)} \beta^2$ and $\beta \approx t_\text{C}/\epsilon$ from the inter-dot tunnel coupling $t_\text{C}$ (full gap). $\epsilon$ is calibrated with the measured lever arm parallel to the QD\ $\leftrightarrow$\ lead transition ($32.4 \ueVpmV$), and the QD loading rate $\varGamma_{(2,1)\leftarrow(1,1)} = 2.56 \MHz$ is measured independently. The fit yields $t_\text{C} = 0.54 \ueV$. This value is consistent with independent measurements through diabadic/adiabatic passage experiments.}
   \label{fig:relaxvsdetuning}
\end{figure}

Using long single-shot readout traces, we can measure the state relaxation and excitation times $T_\text{rel}$ and $T_\text{exc}$ in the different regions \cite{thalakulam2011}. The procedure is similar to that described in the Supplementary information section ``Estimating the relaxation time''. As previously discussed, these times set an upper bound for how fast one should measure to achieve high fidelity. We show the results in \tabref{tab:readoutcomp} and \figref{fig:relaxvsdetuning}. The results clearly show the benefits of the modified reverse ELR relaxation mechanism, which increases $T_\text{rel}$ by a factor of over 100. 

As mentioned in section \ref{sec:fidelitymetric}, both the direct and reverse variations suffer from mapping errors. In the case of the direct ELR, a slow $(2,0) \rightarrow (1,0)$ transition can look like a fast $T_0$ decay and cause an error. In the case of the reverse ELR, a branch in the process ladder results in a mapping error for the triplet of
\begin{align}
	e_{\text{map},T} = \frac{\varGamma_{(2,0)S\leftarrow(1,1)T_0}}{\varGamma_{(2,0)S\leftarrow(1,1)T_0}+\varGamma_{(2,1)\leftarrow(1,1)T_0}} .
	\label{eq:eMapTriplet}
\end{align}
In the present experiment, this mapping error is of the order of $0.1\pc$, larger than the mapping error of order of $0.01\pc$ for the direct ELR (see \tabref{tab:readoutcomp}). However, further optimization of the system parameters (e.g., the loading rates) could bring this error source to a negligible level.

The effect of the charge enhancement can also be seen in the amplitude of the signal, while the noise remains the same. 
\begin{table}
\caption{Comparison of readout parameters between the PSBR and the two ELR variations for donor 1. Parameters are measured using methods described in sections \ref{sec:fidelitymetric}, \ref{sec:readoutcomp}, \ref{sec:highFidelity} and the Supplementary information. The effect of the different relaxation mechanism of the reverse ELR can clearly be seen as it makes the relaxation time over 100 times longer than the PSBR and over 30 times longer than the direct ELR. However, it is important to account for the branching ratio error contribution to $\bar e_\text{map}$, which for these specific QD loading and relaxation rates can limit the benefit of the larger and longer signal. Optimizing loading rates could reduce this to a negligible level. While the direct ELR improves the lifetime only moderately ($\sim 3$ times more than PSBR) compared to the reverse ELR, it can still be very useful by reducing the time required for the readout due to the larger signal and typically has a smaller $\bar e_\text{map}$ due to the unloading rates being faster than the loading rates.}
\begin{center}
\begin{tabular}{l c c c}
	\hline \hline 
	\bf Readout & $\boldsymbol{T_\text{rel}}$  & $\boldsymbol{\bar e_\text{map}}$ & \bf Signal\\ \hline
	PSBR & $300\pm80 \us$  & $0\pc$ & $163 \pA$ \\
	direct ELR & $940\pm60\us$  & $0.007\pc$ & $228 \pA$ \\ 
	reverse ELR & $31\pm2 \ms$  & $0.07\pc$ & $220 \pA$ \\ 
	\hline \hline
\end{tabular}
\end{center}
\label{tab:readoutcomp}
\end{table}

\subsection{Charge-admixture relaxation mechanism}
\label{sec:relaxmech}

In \secref{sec:conditionsELR}, we introduced a charge-admixture relaxation mechanism. In this section, we present evidence for this mechanism.
The effect is most clearly observed in the reverse ELR $T_\text{rel}$ data, see \figref{fig:relaxvsdetuning}(b). We find that it fits a simple relaxation model based on the hybridization between the $(2,0)S$ and $(1,1)S$ states and correctly predicts $t_\text{C}$ based on independently measured parameters. 
Next, according to the schematic of \figref{fig:mechanism}(a), the edges of the readout window should align with those in the PSB region. Experimentally, we find that these are offset a certain amount towards the $(1,1)$ region in the direct variation (\figref{fig:mechanism}(e)) and towards the $(2,0)$ region in the reverse variation (\figref{fig:mechanism}(f)). The offset increases as the measurement time is made longer. 
Finally, we typically observe that the charge latching lifetime during the readout is several orders of magnitude shorter (in this case, ms) than the D ${\leftrightarrow}$ lead tunnel rate far from the anti-crossing (in this case, s). 
Since the enhanced relaxation occurs near the inter-dot degeneracy line (e.g.\ in the PSB readout window), it can be unnoticed in large charge-stability diagrams.

\subsection{High-fidelity single-shot readout} 
\label{sec:highFidelity}

\begin{figure}
   \centering
   \includegraphics[scale=\figscale]{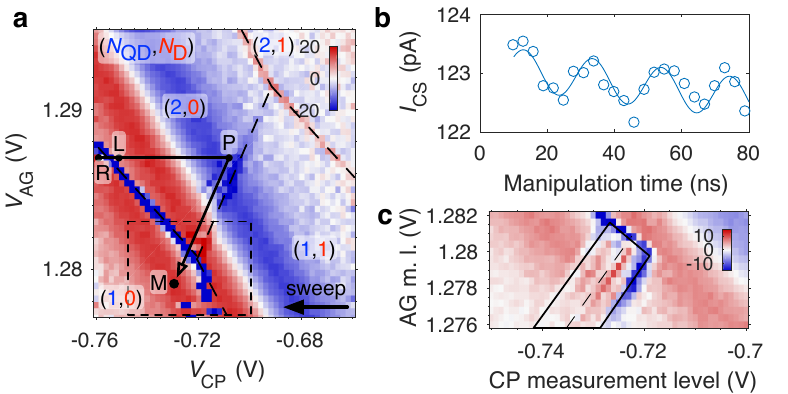} 
   \caption{\figletter{a} Anti-crossing of donor 2 (no pulse applied) and pulse sequence used to demonstrate the high-fidelity direct ELR. Colour scale: d$I_\text{CS}$/d$V_\text{CP}$ (a.~u.).    \figletter{b} This QD-D anti-crossing can produce low-visibility hyperfine-driven coherent rotations as in \refref{harvey-collard2017a}. The visibility is limited by the low bandwidth of the pulse compared with the $\tC$ and effective magnetic field difference $\DBz$.     \figletter{c} An averaged measurement technique is used to image the edges of the enhancement region by varying the location of the measurement point.}
   \label{fig:highfidelitypulse}
\end{figure}

We now demonstrate that the ELR can achieve higher fidelities than the PSB readout using optimized device parameters and a different donor, called donor 2. The pulse sequence is shown in \figref{fig:highfidelitypulse}(a).
As described previously, an averaged measurement technique can be used to image the edges of the readout window, see \figref{fig:highfidelitypulse}(c). Using methods described in \refref{harvey-collard2017a}, we show that this anti-crossing can produce hyperfine-driven coherent rotations between the $S$ and $T_0$ states, see \figref{fig:highfidelitypulse}(b). The visibility of the rotations is low due to experimental bandwidth limits in the pulsing lines ($\sim 10\ns$ resistance-capacity (RC) constant), which prevented us from reaching the rapid adiabatic passage regime. However, these rotations are presented solely as a justification that the parameter regime chosen for the readout demonstration is appropriate for a ST qubit. 

To characterize $\bar F$, we perform an experiment where we prepare singlets and triplets at random and analyze the process chains for the PSB readout and the direct ELR. Specifically, we first look at errors occurring once the M point is reached. We use this experiment and various others to characterize the parameters of the system (e.g.\ tunnel rates, relaxation times) and use this information to calculate additional errors $\bar e_\text{map}$. 
\begin{figure}
   \centering
   \includegraphics[scale=\figscale]{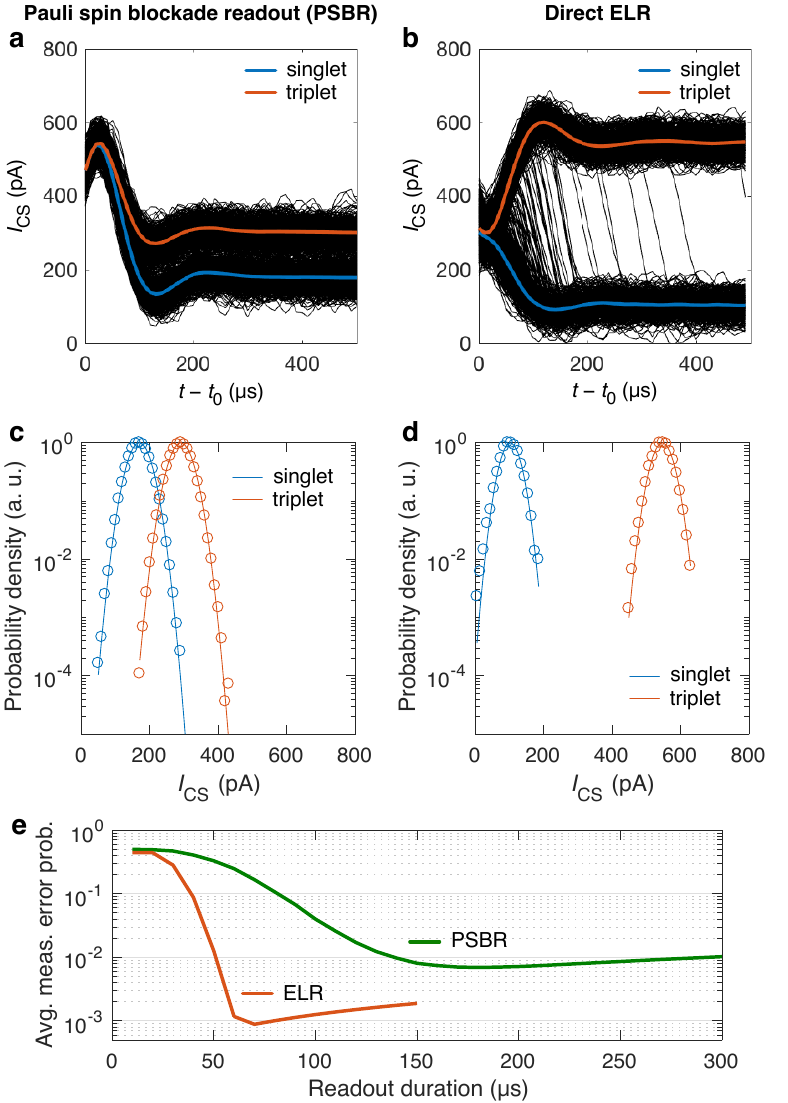} 
   \caption{\figletter{a-b} Single-shot time traces for the PSB readout and ELR of donor 2. The enhancement in signal is clearly visible for the ELR case. \figletter{c-d} Probability density of finding a certain CS current after the signal is stabilized for a $10 \us$ time bin. It shows the enhancement in signal while the noise magnitude stays the same and is well modeled by a Gaussian fit with standard deviation $26.8 \pA$. \figletter{e} The error probability $\bar e_\text{meas}$ first decreases as the readout duration is increased due to the reduced effect of noise in the state estimation. If the duration is too long, errors from the relaxation/excitation become the dominant source. The ELR shows a factor of 8 enhancement in charge measurement error, 2.3 reduction in measurement time, and 3.7 improvement in signal. The limiting factor to the detection time is the limited bandwidth of the system which introduces a significant rise time to the signal, despite the large signal-to-noise ratio of the ELR for $10 \us$ time bins. This suggests large improvements are still possible.}
   \label{fig:highfidelity}
\end{figure}

\subsubsection{Measurement errors}

We show single-shot time traces for the two readouts in \figref{fig:highfidelity}(a-b). They are acquired using the same nominal conditions under a short period of time to allow the best comparison. We define a time $t_0$ where point M is reached. An approximately $90 \us$ rise time can be seen in the traces. This is because the readout line has an RC filter that delays the response. As a result, the signal at the beginning of the cycle has memory of the previous one (not shown). This is not ideal and will be addressed in future experiments, although it does not impact our analysis for this particular experiment. We note that we have subtracted a large systematic $820 \Hz$ signal created by the turbo pump on the refrigerator. It is possible to do so because the amplitude and phase are consistent over time, making them predictable in real time using e.g.\ a Kalman filter \cite{grewal2002}. Using post selection, we can determine the average signal for each of the singlet and triplet signals. These are shown as red and blue thick lines in \figref{fig:highfidelity}(a-b). The remaining noise on these traces is well modeled by a Gaussian fit, as shown in \figref{fig:highfidelity}(c-d). We observe that the noise is the same for the two readouts, but the signal amplitude is $3.7$ times bigger for the ELR case. Using traces of longer duration, we can extract the relaxation rates for the excited and metastable states. We then plot the error probability as a function of the time needed to determine the state associated with the signal in \figref{fig:highfidelity}(e). The details of the single-shot processing are given in the Supplementary information section ``Processing of single-shot charge readout traces''. The error probability initially goes down as more time allows a more accurate determination. Moreover, it takes a non-negligible amount of time for the two signals to separate from one another. At longer times, relaxation/excitation events become dominant and limit the error probability. For the PSB readout, we find $T_\text{rel} = 15\pm3 \ms$ and $T_\text{exc} \sim 300\pm200 \ms$. The excitation events are rare but not negligible for our low error levels. The excited state population is greater than what is expected from the electronic temperature. This could indicate heating from the pulses or excitations driven by the proximity of the anti-crossing. For the direct ELR, we find $T_\text{rel} = 40\pm4 \ms$ and $T_\text{exc} \sim 7 \s$ (from the rare events available). The excitation is found to be negligible for the enhanced case, which is consistent with the larger energy gap that separates the states. The reduction in charge excitation is yet another benefit of the ELR. The time required for determining the charge $T_\text{meas}$ is reduced from $150 \us$ in the PSB case to $65 \us$ in the ELR case, limited by the signal rise time. This 2.3 times improvement is possible because of the larger signal. This in turn reduces the $\bar e_\text{meas}$ from $0.7\pm0.1\pc$ to $0.088\pm0.008\pc$, a factor of 8 improvement.

\subsubsection{Mapping errors}

Mapping errors are in general summarized by the various arrows of \figref{fig:mechanism}(c-d) that can lead to incorrect inference of the spin state. A complete assessment of these processes needs to account for relaxation, excitation, tunnel rates, and pulse sequence parameters/trajectory. The various processes contributing to $\bar e_\text{map}$ are detailed in the Supplementary information section ``Mapping error for high-fidelity result of donor 2''. We find $\bar e_\text{map} = 0.048\pc$. This comes mostly from a $20\us$ ramp that is used to make sure that the pulse trajectory is carefully followed and is necessary because of the low bandwidth of the AG gate. This could be improved and virtually eliminated with system optimizations. 

Combining these errors together, we find an average error probability of $\bar e_\text{PSBR} = 0.7\pc$ and $\bar e_\text{ELR} = 0.136\pc$. The results are summarized in \tabref{tab:highfidelity}.
\begin{table}
\caption{Comparison of readout parameters between the PSBR and the direct ELR for donor 2. The reverse ELR was not performing well in this case because the metastable lifetime was short. We suspect this could be caused by the larger inter-dot tunnel coupling. Despite this, the direct ELR worked well. We repeatedly see the charge latching get softer as the total electron number goes up, like in \citet{yang2014a}. We hypothesize that this can be favorable to the direct ELR in the donor 2 case since it involves states with fewer electrons.}
\begin{center}
\begin{tabular}{l c c c c c c}
	\hline \hline 
	\bf Readout & $\boldsymbol{T_\text{rel}}$ & $\boldsymbol{T_\text{meas}}$ & \bf Signal & $\boldsymbol{\bar e_\text{map}}$ & $\boldsymbol{\bar e_\text{meas}}$ & $\boldsymbol{\bar e_\text{tot}}$\\ \hline
	PSBR & $15\ms$ & $150\us$ & $121 \pA$ & $0\pc$ & $0.7\pc$ & $0.7\pc$ \\
	direct ELR & $40\ms$ & $65\us$ & $444 \pA$ & $0.048\pc$ & $0.088\pc$ & $0.136\pc$\\ 
	\hline \hline
\end{tabular}
\end{center}
\label{tab:highfidelity}
\end{table}

\subsubsection{High-fidelity singlet preparation and readout}

As a complementary test, we measure the state preparation and measurement errors for pure singlets. High-fidelity singlet states are prepared by loading the $(2,0)$ ground state, then measured using the pulse sequence described in \figref{fig:highfidelitypulse}a. We find very few triplet counts, corresponding to a small $e_S < 0.1 \pc$ over $10^5$ cycles. This result is consistent with the stated readout fidelity. 

\section{Discussion}

\subsection{What is the best readout?}

It is worth noting that the best readout to use depends on the specific details of the system. The variables to consider are the lifetime enhancement and the mapping error overhead. For instance, in GaAs the PSB relaxation time is typically $\sim 10 \us$ and fast radio-frequency readouts can measure in $\sim 1 \us$ \cite{barthel2012}. Such a case could clearly benefit from the lifetime enhancement of the reverse ELR, even at the expense of extra mapping errors. In Si, this relaxation time can be tens of milliseconds. In such a case the better option can be either direct or reverse ELR, depending on the mapping errors and the degree of lifetime enhancement. In this work, the direct ELR performed better for donor 2. However, taking advantage of the parametric lifetime enhancement of the reverse ELR could further improve the fidelity. With really fast charge readout capabilities, the mapping errors can easily become a limiting factor.

The reverse ELR changes the relaxation mechanism from a spin one to a charge-metastable one. In cases like donor 1, this can lead to dramatic improvements of the lifetime of the signal, as is proved for the first time in this work. We also have identified that the factor limiting this charge lifetime is charge hybridization (see section \ref{sec:relaxmech}). This can also lead to the suppression of the charge lifetime in the spin blockade region, particularly for strong tunnel couplings (see \tabref{tab:highfidelity} caption).

\subsection{How much signal enhancement?}

The degree of signal enhancement depends on the system geometry. In this work, we focused on cases where the inter-dot transition is visible, which enables the performance comparison. We have observed improvements of $1.4$ and $3.7$ times for our two donors. Importantly, we have also used the ELR on both donor-dot and double-QD qubits when the inter-dot signal vanishes due to the alignment of the charge dipole. In those cases, the signal (and fidelity) enhancement is very large, because readout is otherwise not possible using the traditional PSBR.

\subsection{Other ELR variations?}

There are other variations of the ELR. In this work, the two-electron state has the electrons on the QD that has direct access to a charge reservoir. Interchanging which QD has the two-electron state is expected to also interchange the lifetime properties of the direct and reverse ELR variations.

\subsection{Conclusion}

In summary, we have demonstrated that the enhanced latching mechanism described in this work can achieve high-fidelity single-shot readout for a spin qubit. The cost in fidelity due to the charge enhancement processes can be optimized such that the overall fidelity of the process is improved. This is done through direct comparison between the conventional PSBR and the ELR. We demonstrate a readout fidelity $> 99.86\pc$ in $65 \us$, the highest reported so far for spin blockade. Total readout time is limited by the readout circuit and could be reduced using cryogenic amplification techniques.

A central contribution of this work is to elucidate critical microscopic mechanisms that contribute to errors in the ELR. Therefore, it provides guidance to improve the fidelity beyond $99.9\pc$. In particular, we discuss two variations of the ELR that each have benefits and tradeoffs. Both variations improve the fidelity by improving the signal amplitude. In addition, one is shown to replace the spin-relaxation mechanism by a charge-metastable one. This can improve the signal lifetime by a factor over 100. The metastable charge lifetime is limited by the hybridization between the $(1,1)$ and $(2,0)$ states that occurs near the anti-crossing due to the tunnel coupling.

Finally, we also highlight that the enhancement process also relaxes restrictions on the CS layout considerably. Conventional PSBR requires careful alignment of the CS with respect to the $(2,0)-(1,1)$ dipole transition. This alignment can be particularly challenging in the donor-based qubit system demonstrated in this work. The benefits are applicable to any QD qubits as well.

\section{Methods}

\subsection{Device}
Electrons are confined in a 2D electron gas at the interface between an epitaxial enriched \ce{^{28}Si} layer with $500 \text{ ppm}$ residual \ce{^{29}Si} and a $35\nm$ gate oxide. Highly n-doped poly-silicon gates are patterned on top of the gate oxide using low pressure chemical vapor deposition and plasma etching \cite{tracy2013}. The gate structure is shown in \figref{fig:experiment}(a). These are used to accumulate electrons in an enhancement mode by applying a positive voltage or deplete electrons with negative voltages. Phosphorus donors are implanted in a PMMA resist window that overlaps with the AG gate on both sides of both wires, but only the donors near the red dot in \figref{fig:experiment}(a-b) are important for this work. 
\subsection{Charge sensing and measurements}
Experiments are performed in $200 \mT$ (donor 1 data) and $300 \mT$ (donor 2 data) in-plane magnetic fields. The measured electron temperature is $207 \mK$. For charge stability diagrams, the current through the CS $I_\text{CS}$ is measured at $0 \V$ DC bias with a lock-in measurement using an AC excitation of $40 \uVrms$ (donor 1 data) and $100 \uVrms$ (donor 2 data) at $454 \Hz$. The derivative of the CS current is taken to show the sharp steps indicating charge transitions in the QD-D system. The oscillating background in charge stability plots is the Coulomb peaks of the CS. The ST splittings were measured to be $94 \ueV$ (donor 1 data) and $222 \ueV$ (donor 2 data).
\subsection{Pulsing and single-shot} \label{sec:methodsPulsing}
For single-shot measurements, the CS is DC-biased with voltages of $100 \uV$ (donor 1 data) and $60 \uV$ (donor 2 data). For the donor 2 data the CS was tuned to have a very narrow and conductive peak to maximize the response. The pulses are applied to the device using a Tektronix AWG7122C arbitrary waveform generator. The pulses are applied through a room-temperature RC bias tee. Waveforms are generated so that all target points are fixed relative to the charge stability diagram except for some parameters that are swept (e.g.\ measurement point location). The single-shot current traces are filtered through an RC cryogenic filter, amplified using a DL 1211 current pre-amplifier, and measured using a Keysight DSO-X 4104A oscilloscope.
\subsection{Tunnel rates}
The QD ${\leftrightarrow}$ lead loading/unloading rates are $2.56\MHz/22.6\MHz$ for donor 1 and $14\kHz/400\kHz$ for donor 2. 
The full-gap QD-D tunnel couplings $\tC$ are $\sim0.5\ueV = h\times120\MHz$ for donor 1 and $\gtrsim 20\ueV = h\times4.8\GHz$ for donor 2 (here $h$ is the Planck constant). The D ${\leftrightarrow}$ lead direct tunnel rates are $> 1 \s$ for both donors.

\section*{Acknowledgements}
The authors would like to thank Stephen Carr for valuable help regarding this work. P.H.-C.\ acknowledges funding from Canada's National Science and Engineering Research Council (NSERC). W.A.C.\ acknowledges funding from NSERC and the Canadian Institute for Advanced Research (CIFAR).
This work was performed, in part, at the Center for Integrated Nanotechnologies, an Office of Science User Facility operated for the U.S.\ Department of Energy (DOE) Office of Science. Sandia National Laboratories is a multimission laboratory managed and operated by National Technology and Engineering Solutions of Sandia, LLC, a wholly owned subsidiary of Honeywell International, Inc., for the DOE's National Nuclear Security Administration under contract DE-NA0003525.

\section*{Author contributions}
P.H.-C.\ designed the readout and performed the experiments. 
B.D.\ and W.A.C.\ performed the single-shot fidelity analysis.  
P.H.-C., B.D.\ and W.A.C.\ found the dominant metastable relaxation mechanism for the readout. 
P.H.-C., B.D., W.A.C., M.S.C.\ and M.P.-L.\ analyzed and discussed central results throughout the project, including designing experiments and models. 
M.R.\ performed supporting measurements on similar control samples that establish repeatability of many observations in this work.  
N.T.J., P.H.-C.\ and M.R.\ modelled key elements of the device structure providing critical insights.  
J.D., T.P., G.A.T.E.\ and M.S.C.\ designed process flow, fabricated devices and designed/characterized the $^{28}$Si material growth for this work.
J.R.W.\ performed critical nanolithography steps. 
M.L.\ supplied critical laboratory set-up for the work. 
M.S.C.\ supervised the combined effort including coordinating fabrication, measurement and modelling. 
P.H.-C., B.D., M.S.C.\ and M.P.-L.\ wrote the manuscript with input from all co-authors.

\section*{Supplementary information}
Supplementary information is provided with this paper.

\bibliographystyle{apsrev4-1-title} 
\bibliography{/Users/Patrick/Documents/Papers/PHC}


\newcommand{\RefFigHighfi}{\figref{fig:highfidelity}}
\newcommand{\RefSecHighfi}{\secref{sec:highFidelity}}
\newcommand{\RefSecMethodspulsing}{\secref{sec:methodsPulsing}}

\makeatletter
\renewcommand\thesection{\mbox{S\arabic{section}}}
\makeatother
\setcounter{section}{0}     

\newcounter{supfigure} \setcounter{supfigure}{0} 
\makeatletter
\renewcommand\thefigure{\mbox{S\arabic{supfigure}}}
\makeatother

\newcounter{suptable} \setcounter{suptable}{0} 
\makeatletter
\renewcommand\thetable{\mbox{S\arabic{suptable}}}
\makeatother

\newenvironment{supfigure}[1][]{\begin{figure}[#1]\addtocounter{supfigure}{1}}{\end{figure}}
\newenvironment{supfigure*}[1][]{\begin{figure*}[#1]\addtocounter{supfigure}{1}}{\end{figure*}}
\newenvironment{suptable}[1][]{\begin{table}[#1]\addtocounter{suptable}{1}}{\end{table}}
\newenvironment{suptable*}[1][]{\begin{table*}[#1]\addtocounter{suptable}{1}}{\end{table*}}

\makeatletter
\renewcommand\theequation{S\arabic{equation}}
\makeatother
\setcounter{equation}{0}

\clearpage
\newpage
\onecolumngrid
{\centering
\large\textbf
{Supplementary information for: \\ \mytitle} \\ \rule{0pt}{12pt}
}
\twocolumngrid


\section{Error composition formula}

In this section, we derive formulas for the composition of classical and independent error processes in series. We assume two classical states $S$ and $T$. Let $e_S$ and $e_T$ be the probability that $S$ flips to $T$, and $T$ flips to $S$, respectively. We define an average error probability $\bar e = (e_S+e_T)/2$. Let $\bar e' = (e'_S+e'_T)/2$ be the error probabilities for a second process occurring in series after the first process. The total error of these two processes in series is such that $e_S''=v'e_S+e_S'$, $e_T''=v'e_T+e_T'$, with $v'=1-e_S'-e_T'$. The total average error probability is then $\bar e'' = \bar e' + \bar e -2 \bar e' \bar e$. If $\bar e', \bar e \ll 1$, the second order term can be neglected and the average errors simply add, $\bar e'' \approx \bar e' + \bar e$.

\section{Mapping error for high-fidelity result of donor 2} \label{sec:mappingErrorSupp}

The visibility of the ST rotations allows to estimate the inter-dot $\tC$ to be $\gtrsim 20 \ueV$ (full gap at degeneracy point).

Next, we look at the unload rate of the $(2,0)S$ state, $\varGamma_{(1,0)\leftarrow(2,0)S}$. Using a detailed model for excited state spectroscopy, we extract a lower bound for this rate and find $1/\varGamma_{(1,0)\leftarrow(2,0)S} < 2.5 \us$. We estimate that for $t-t_0 = 13 \us$, and knowing that when $t_0$ is reached the system has already spent $10\us$ in the $(1,0)$ region due to the ramp, the probability for a singlet to be read out as a decaying triplet is $<1\times10^{-4}$. This adds a rather small delay to the charge signal in a $<1\times10^{-4}$ number of cases (for singlets only) compared to other error sources. It is therefore not a significant error source. 

The possible decay of triplets before $t_0$ can also be a source of errors. Those decaying after are accounted for in the measurement error $\bar e_\text{meas}$. For this experiment we use a fairly long $20 \us$ ramp to go from P to M. This is done to make sure that the pulse trajectory is carefully followed and is necessary because of the low bandwidth of the AG gate. This could be improved and virtually eliminated with system optimizations. The system spends $10\us$ in the PSB region before crossing over to the enhancement region. This allows for some triplets to relax to singlets and some singlets to be excited to triplets before the enhancement. Considering the relaxation/excitation rates found in the PSB readout, this process adds a $e_T = 0.67\times10^{-3}$ and $e_S = 0.033\times10^{-3}$ error probability before the enhancement region. In the enhancement region, the triplet decay probability is estimated using the decay rate in the enhanced region, and is $e_T = 0.25\times10^{-3}$. The singlet excitation before $(2,0)S$ is mapped to $(1,0)$ is given by a branching ratio between $\varGamma_{(1,0)\leftarrow(2,0)S}$ and $\varGamma_{(1,1)T_0\leftarrow(2,0)S}$ and is $e_S = 0.01\times10^{-3}$. 

Finally, we look at thermal excitation of $(1,1)T_0$ into $(2,0)T_0$ followed by an immediate decay into $(1,0)$. We suspect from the long lifetime measured that this is not an issue. We nevertheless estimate through detailed balance that the excitation rate is $0.1\times10^{-3} \times \varGamma_{(1,1)T_0\leftarrow(2,0)T_0}$. Though we do not measure this rate, if we assume it is governed by a similar mechanism as PSB relaxation, it should be several seconds and therefore negligible.

The total average mapping error is therefore $0.048\pc$.

\section{Processing of single-shot charge readout traces} \label{sec:suppProcessing}

\subsection{Removing unwanted signal from turbo pump}

The acquired single-shot readout traces were distorted by a periodic signal of frequency $\sim 820\,\textrm{Hz}$ generated by a turbo pump. This signal complicates the interpretation of the readout statistics. While it would be ideal to directly isolate the readout chain from this spurious signal, here we show that we can demonstrate single-shot readout capability even when it is present.

Since the signal is periodic, we can expand it in a Fourier series:
\begin{align}
	s_{\textrm{TP}}(t) = \sum_{k=1}^{\infty} A_k \cos\left[2\pi k f t + \phi_k\right] \label{eq:turboFitForm}
\end{align}
where $f\approx 820\,\textrm{Hz}$ is the pump signal frequency and where $A_k$ and $\phi_k$ are the amplitudes and phases of the pump signal harmonics. We wish to estimate the parameters $f$, $\left\{A_k\right\}$, and $\left\{\phi_k\right\}$.

Since the pump signal period is smaller than but comparable to the duty cycle of the single-shot readout, we need to estimate the parameters in Eq.\ \eqref{eq:turboFitForm} over many such cycles while tracking the phase of the oscillations. Figure \ref{fig:figC1}a shows a subset of $\sim 18$ single-shot readout cycles for the Pauli spin blockade readout of \RefSecHighfi{}. To isolate the effect of the pump signal from the readout signal, we postselect time intervals within each cycle where the charge state has relaxed and where the initial transient response of the charge sensor (seen in Figs.\ \ref{fig:figC2}a-b) has disappeared. More precisely, we first select cycles for which the time-averaged current over a pump period and the maximum charge sensor current on the period are both below a threshold (chosen intermediate between the two charge sensor readout signals). For each of these cycles, we keep a short time interval at the end of the cycle where the initial transient response of the charge sensor has disappeared. These time intervals are shown in \figref{fig:figC1}a. We then subtract the average charge signal of the selected time intervals to obtain a pump signal of the form in Eq.\ \eqref{eq:turboFitForm}, as illustrated in \figref{fig:figC1}b. We perform a similar analysis for the enhanced latching readout.

To maintain real-time single-shot readout capability, it is desirable to have the ability to remove the pump signal in real time. We show that this is possible by fitting the extracted pump signal to Eq.\ \eqref{eq:turboFitForm} over a large number of cycles (typically $\gtrsim 10^3$) using an extended Kalman filter. The extended Kalman filter performs a least squares fit to Eq.\ \eqref{eq:turboFitForm} in real-time, demonstrating the ability to estimate and remove the turbo pump signal on-the-fly. The fit accurately reproduces the measured signal, as can be seen in \figref{fig:figC1}b. We find that it is sufficient to keep the first 5 terms in Eq.\ \eqref{eq:turboFitForm}. We also find that the pump signal parameters do not drift substantially over the acquisition period.

Finally, we subtract the fitted signal from the entire readout sequence. The resulting single-shot readout traces are those shown in, e.g., \RefFigHighfi{}(a-b).
\begin{supfigure}
\centering
\includegraphics[width=8.25cm]{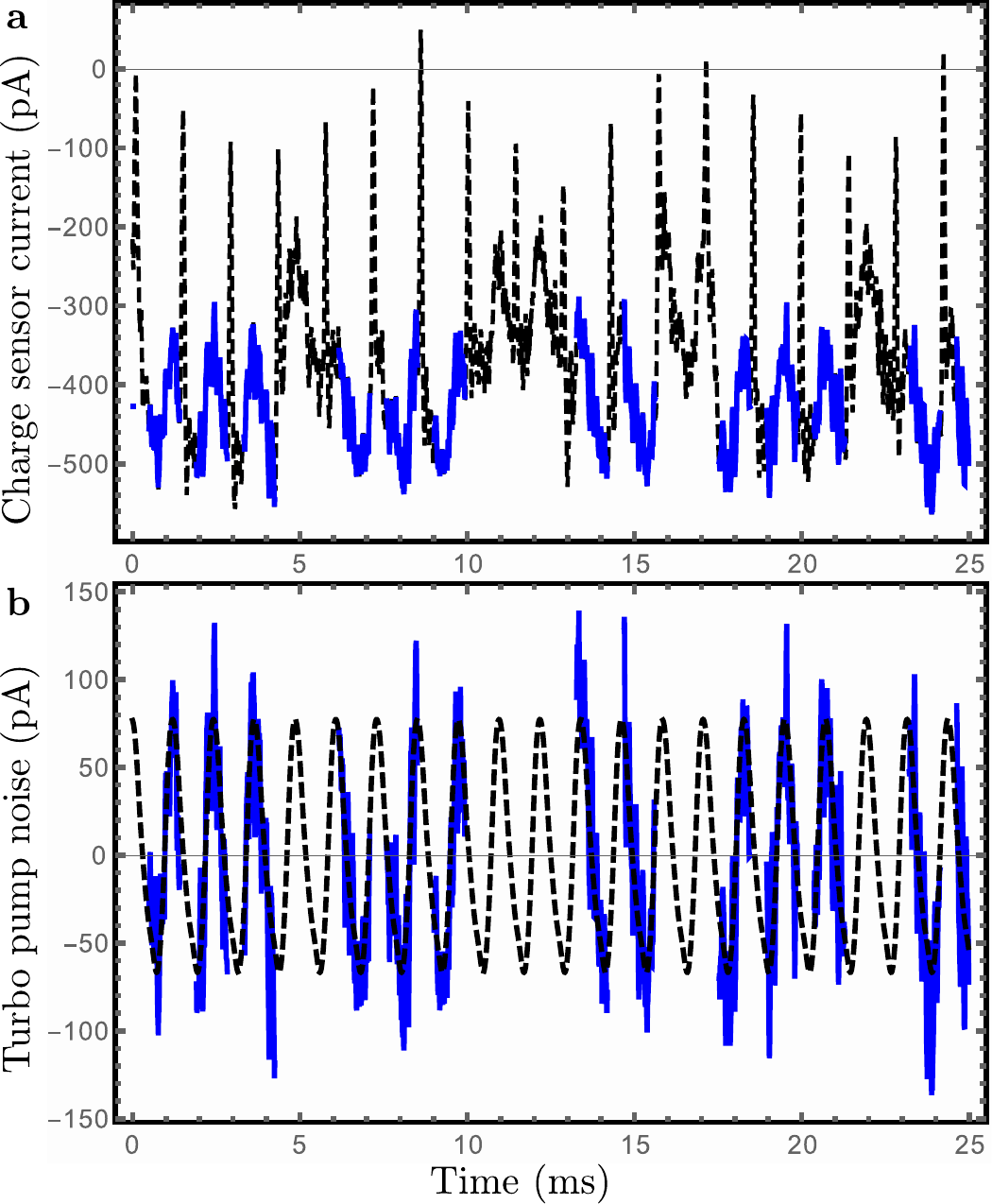}
\centering
\caption{\figletter{a} A few ($\sim 18$) single-shot readout cycles (dashed black line). For each cycle, we extract a time interval where the charge state has relaxed and where the initial transient response of the charge sensor has disappeared (solid blue line).     \figletter{b} Resulting measured turbo pump signal (solid blue line) and fit obtained using the Kalman filter (dashed black line). \label{fig:figC1}}
\end{supfigure}
	
\subsection{Estimating the charge sensor signal and noise \label{sec:signalAndNoise}}

Estimating the charge readout fidelity requires a precise understanding of the charge sensor signal and noise conditioned on the initial charge state. To characterize the signal and noise for each state, we postselect the single-shot readout cycles for which no relaxation or excitation occurs. For this, we select readout cycles for which the maximum (minimum) charge sensor current over the time interval where the two charge signals can be discriminated is below (above) a threshold (chosen to be intermediate between the two charge sensor readout signals). The two sets of postselected traces are illustrated in \figref{fig:figC1}a-b for the Pauli spin blockade readout of \RefSecHighfi{}. We perform the same analysis for the enhanced latching readout.

From the postselected traces, we can immediately extract the mean sensor signal, as shown in \figref{fig:figC2}a-b. We then verify that the noise has Gaussian statistics by making a histogram of the fluctuations of the charge sensor current around the mean (see \RefFigHighfi{}(c-d) in the main text). Gaussian noise is completely characterized by its covariance. We thus calculate the covariance matrix of the noise for the two possible charge states. As illustrated in \figref{fig:figC2}c-d, the covariance matrices capture non-stationary features of the noise (greater noise during the finite signal rise time) and non-Markovian features of the noise (oscillations along the off-diagonal of the covariance matrix). We account for all these features in our charge readout fidelity estimate (see \secref{sec:fidelityBound}).
\begin{supfigure}
\centering
\includegraphics[width=8.25cm]{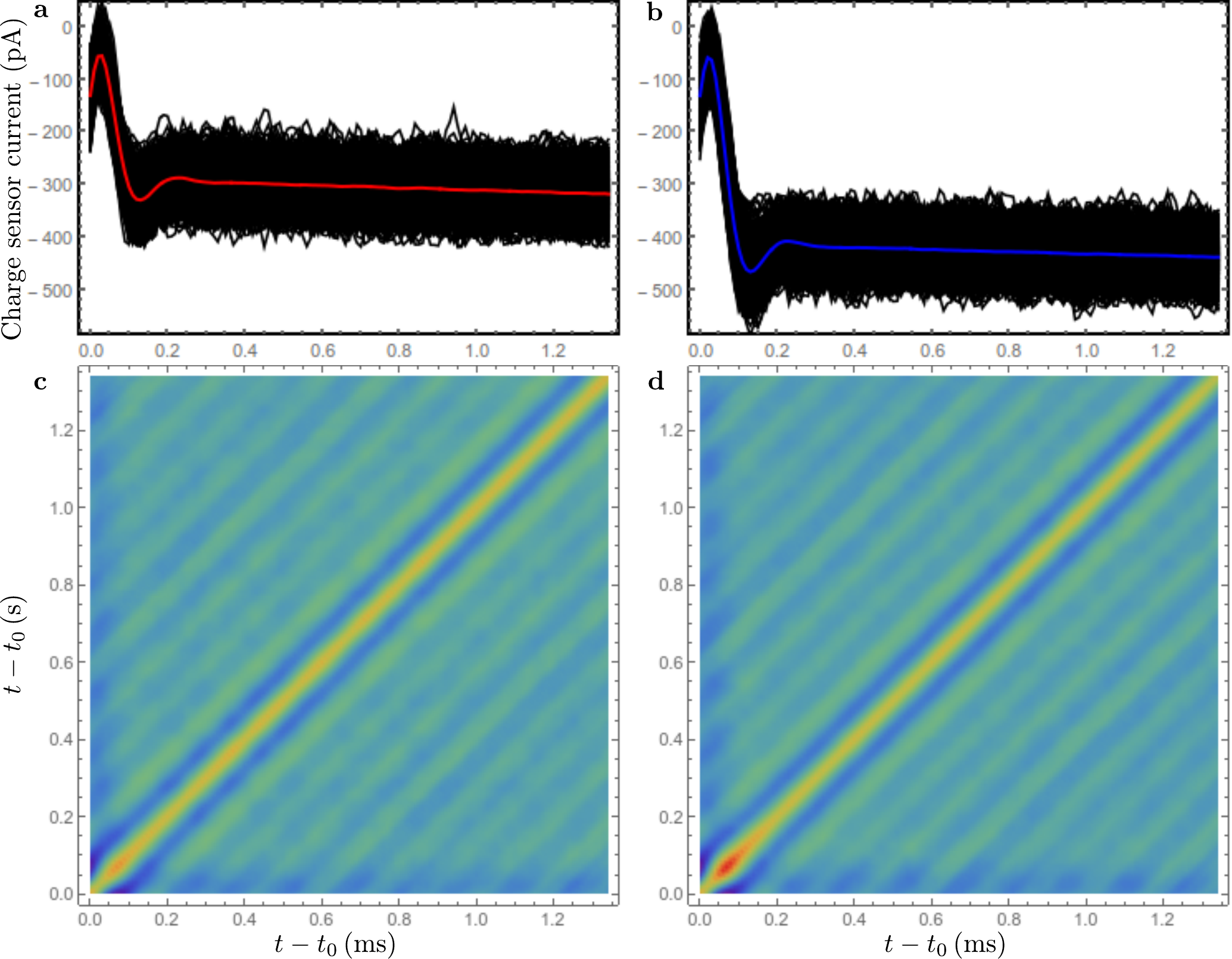}
\centering
\caption{\figletter{a-b} Single-shot traces postselected on the absence of relaxation or excitation (solid black) for the Pauli spin blockade readout of \RefSecHighfi{}. We also show the mean signal for the (1,1) charge state (solid red) and the (2,0) charge state (solid blue). The persistent slant in the sensor response is attributed to a bias-tee transient, as discussed in \secref{sec:estimateRelaxationTimes}. \figletter{c-d} Covariance matrices of the noise postselected on the (1,1) charge state, (c), and on the (2,0) charge state, (d). Time is measured from $t_0$, the time of arrival at the measurement point.\label{fig:figC2}}
\end{supfigure}
	
\subsection{Estimating the relaxation time \label{sec:estimateRelaxationTimes}}

The charge readout fidelity is ultimately limited by the finite integration time set by the relaxation and excitation processes. Suppose the charge sensor signals for both possible initial charge states, say $\left|+\right>$ and $\left|-\right>$, are constants $I_+$ and $I_-$.  We can then extract the relaxation rate $\varGamma_{\textrm{rel.}}$ from $\left|+\right>$ to $\left|-\right>$ and the excitation rate $\varGamma_{\textrm{exc.}}$ from $\left|-\right>$ to $\left|+\right>$ by fitting the ensemble average of the single-shot readout traces, $\left<I_{\textrm{CS}}(t)\right>$, to an asymmetric telegraph process:
\begin{align}
\begin{split}
	& \left<I_{\textrm{CS}}(t)\right> = A e^{-\varGamma t} + B ,\\
	& \varGamma = \varGamma_{\textrm{rel.}} + \varGamma_{\textrm{exc.}} ,\\
	& B = \frac{\varGamma_{\textrm{exc.}}}{\varGamma} I_+ + \frac{\varGamma_{\textrm{rel.}}}{\varGamma} I_- .
	\label{eq:telegraphProcessFitForm}
\end{split}
\end{align}
Note that Eq.\ \eqref{eq:telegraphProcessFitForm} assumes that the charge dynamics involves only two states. Depending on the system parameters (such as detuning, tunnel coupling, or donor hyperfine coupling), this may not always be a good approximation. Nevertheless, we verified that our fitting procedure gives results consistent with direct counting of the number of relaxation and excitation events.

\figref{fig:figC3}a shows the measured charge sensor signals in the absence of relaxation or excitation for both charge states (see \secref{sec:signalAndNoise}) for the Pauli spin blockade readout of \RefSecHighfi{} (where $\left|+\right> = (1,1)$ and $\left|-\right> = (2,0)$). The measurement conditions are the same as in \figref{fig:figC2}a-b, but with a single-shot duty-cycle of $30\textrm{ ms}$ instead of $1.4\textrm{ ms}$. The charge sensor signal changes over time, which we attribute to (possibly charge-state dependent) bias-tee transients (see \RefSecMethodspulsing{}). Both transients are fit to an exponential and $I_{\pm}$ are extracted from the long-time behavior. In cases where there is not enough data available (i.e.\ too much statistical noise) to accurately fit to an exponential behavior, the transients are fit to a constant instead. To fit the average single-shot signal to Eq.\ \eqref{eq:telegraphProcessFitForm}, we first rescale $\left<I_{\textrm{CS}}(t)\right>$ to remove the effect of the transients. The corrected average is shown in \figref{fig:figC3}b along with the fit to Eq.\ \eqref{eq:telegraphProcessFitForm}. We perform a similar procedure for the enhanced latching readout.

For the Pauli spin blockade readout, charge excitation is observed in a non-negligible fraction of readout cycles. For the enhanced latching readout, however, there are almost no instances of excitation. Therefore, we leave $B$ as a free fit parameter for the Pauli spin blockade readout and we set $B=I_-$ for the enhanced latching readout. We speculate that the excitation in the Pauli spin blockade readout is generated by environmental electric field fluctuations of unknown origin coupling to the $\textrm{(2,0)}\rightarrow\textrm{(1,1)}$ dipole transition. This excitation mechanism is suppressed in the enhanced latching readout because an electron must first be exchanged with the low-temperature reservoir before such a dipole transition can occur. Charge excitation in the direct enhanced latching readout, for example, first requires the transition from $(1,0)$ to $(2,0)$, followed by a transition from $(2,0)$ to $(1,1)$ before the charge state has a chance to go back to $(1,0)$. This is protected by a $>100 \ueV$ energy gap, much larger than the  $k_\text{B}T_\text{e} = 18 \ueV$ electronic temperature.

The precision on our estimates of $\varGamma_{\textrm{rel.}}$ and $\varGamma_{\textrm{exc.}}$ is limited by the statistical fluctuations in $\left<I_{\textrm{CS}}(t)\right>$ due to the finite number of single-shot traces. We thus estimate the error bars on the fit parameters $\varGamma$, $A$, and $B$ by calculating the Fisher information matrix of the parameters with respect to the assumed telegraph process. Note that since the noise in the telegraph process is correlated on a time scale $\varGamma^{-1}$, the parameter we want to estimate, it is important to account for temporal correlations of the telegraph process in the estimate of the error bar. Not accounting for these correlations underestimates the error bar by an order of magnitude.
	
\begin{supfigure}
\centering
\includegraphics[width=8.25cm]{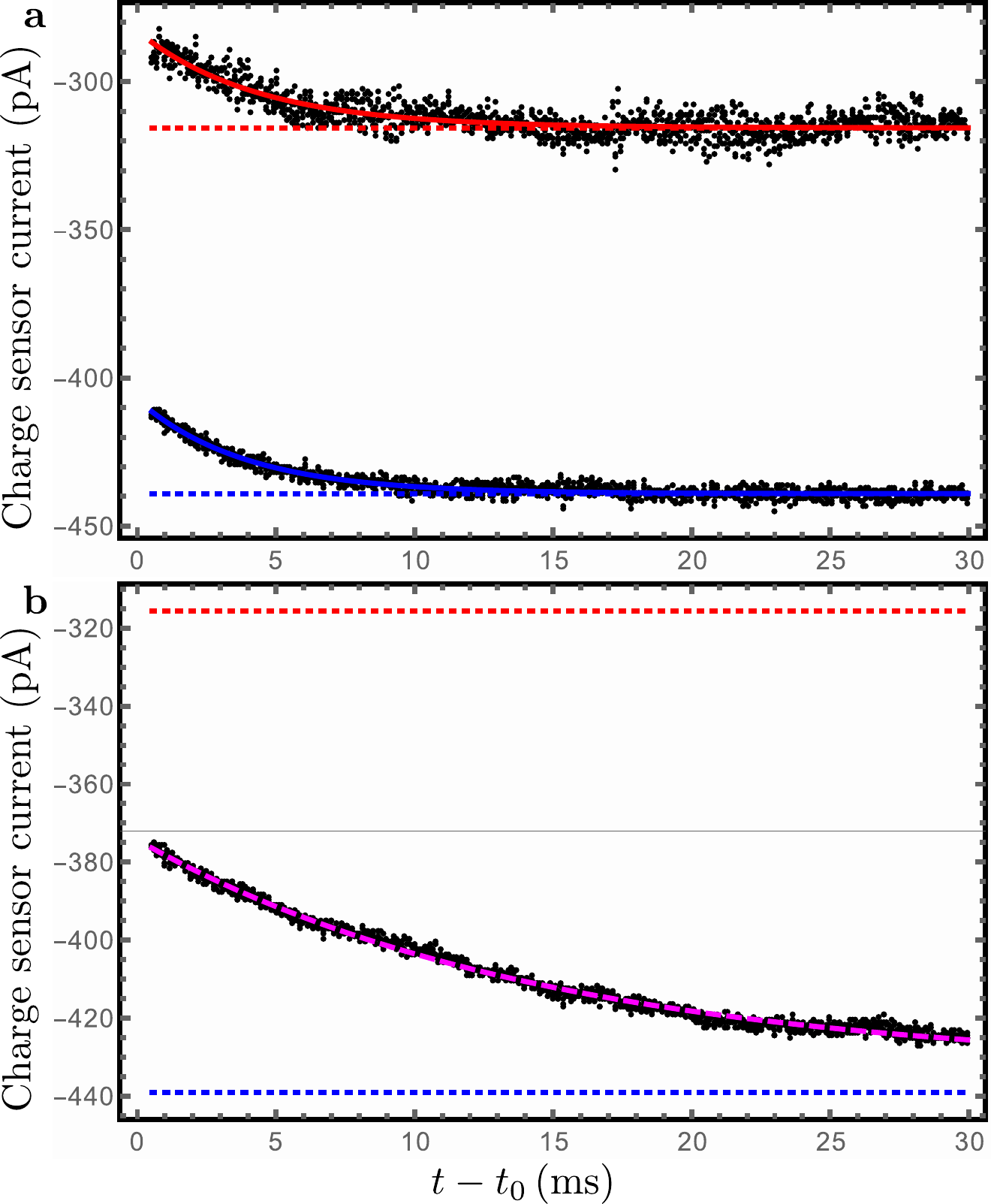}
\centering
\caption{\figletter{a} Observed time dependence of the charge sensor signal for both initial charge states (black dots) for the Pauli spin blockade readout of \RefSecHighfi{}. We fit the signals to extract the transients for $\left|+\right> = (1,1)$ (solid red) and for $\left|-\right> = (2,0)$ (solid blue). The long-time behavior of the fit gives $I_+$ (dotted red) and $I_-$ (dotted blue).      \figletter{b} Average of the single-shot traces (black dots) after correction for the transients. In general, the average decays to some value between $I_+$ (dotted red) and $I_-$ (dotted blue). \label{fig:figC3}}
\end{supfigure}	

\subsection{Charge readout fidelity bound \label{sec:fidelityBound}}

Knowing the signal, noise, and relaxation and excitation rates, we can obtain an upper bound on $\bar{e}_{\textrm{meas.}}$. In the following, we count errors from $t_0$, the time of arrival at the measurement point. Additional mapping errors are discussed in \secref{sec:mappingErrorSupp}.

If the charge state is initially $\left|+\right>$, the error probability is bounded by:
\begin{align}
	e_{\textrm{meas.},+} &= P(\textrm{err.}|+,\textrm{rel.})P(\textrm{rel.}|+) \notag \\
	&\;\;\;\;+ P(\textrm{err.}|+,\textrm{no rel.})P(\textrm{no rel.}|+) \notag ,\\
	e_{\textrm{meas.},+} &< P(\textrm{rel.}|+) + P(\textrm{err.}|+,\textrm{no rel.})P(\textrm{no rel.}|T) \notag ,\\
	e_{\textrm{meas.},+} &< 1 - e^{-\varGamma_{\textrm{rel.}}(t-t_0)} \notag\\
	&\;\;\;\;+ P(\textrm{err.}|+,\textrm{no rel.}) e^{-\varGamma_{\textrm{rel.}} (t-t_0)} \notag \\
	&= e_{\textrm{meas.},+}^B .
	\label{eq:boundErrorTriplet}
\end{align}
Similarly, when the charge state is initially $\left|-\right>$, the error probability is bounded by:
\begin{align}
\begin{split}
	e_{\textrm{meas.},-} &< 1 - e^{-\varGamma_{\textrm{exc.}}(t-t_0)} \\
	&\;\;\;\;+ P(\textrm{err.}|-,\textrm{no exc.}) e^{-\varGamma_{\textrm{exc.}} (t-t_0)} \\
	&= e_{\textrm{meas.},-}^B .
\end{split}\label{eq:boundErrorSinglet}
\end{align}
The average error probability is bounded by:
\begin{align}
\begin{split}
	\bar{e}_{\textrm{meas.}} &= \frac{1}{2} (e_{\textrm{meas.},+} + e_{\textrm{meas.},-}) \\
	& < \frac{1}{2} (e_{\textrm{meas.},+}^B + e_{\textrm{meas.},-}^B) .
\end{split} \label{eq:boundError}
\end{align}
Eqs.~\eqref{eq:boundErrorTriplet}, \eqref{eq:boundErrorSinglet}, and \eqref{eq:boundError} give a bound on the average error probability in the presence of relaxation and excitation, given the error probabilities in the absence of relaxation or excitation, $P(\textrm{err.}|+,\textrm{no rel.})$ and $P(\textrm{err.}|-,\textrm{no exc.})$. In the following, we obtain Monte-Carlo estimates of $P(\textrm{err.}|+,\textrm{no rel.})$ and $P(\textrm{err.}|-,\textrm{no exc.})$.

Let $n = (t-t_0)/\delta t$ be the number of measurement time bins, where $\delta t$ is the sampling time. In the absence of relaxation (excitation), the vector signal $I_n$ for the $\left|+\right>$ ($\left|-\right>$) charge state is a multivariate Gaussian random variable with mean signal $\mu_+$ ($\mu_-$) and covariance matrix $C_+$ ($C_-$). These quantities are directly measured as described in \secref{sec:signalAndNoise}. Formally, we have:
\begin{align}
\begin{split}
	P(I_n|+,\textrm{no rel.}) = \frac{1}{\sqrt{(2\pi)^n |C_+|}} e^{-\frac{1}{2}(I_n-\mu_+)^T C_+^{-1} (I_n-\mu_+)} ,\\
	P(I_n|-,\textrm{no exc.}) = \frac{1}{\sqrt{(2\pi)^n |C_-|}} e^{-\frac{1}{2}(I_n-\mu_-)^T C_-^{-1} (I_n-\mu_-)} .\label{eq:signalDistributions}
\end{split}
\end{align}
We simulate the readout in the absence of relaxation and excitation in the following way. For each readout time and for both initial states, $10^6$ random signals $I_n$ are randomly sampled from the distributions of Eq.\ \eqref{eq:signalDistributions}. Eq.\ \eqref{eq:signalDistributions} is then used to decide which state most likely produced the random signal. Counting the number of errors for each charge state gives $P(\textrm{err.}|+,\textrm{no rel.})$ and $P(\textrm{err.}|-,\textrm{no exc.})$. An upper bound on the error rate is then obtained using Eqs.\ \eqref{eq:boundErrorTriplet}, \eqref{eq:boundErrorSinglet}, and \eqref{eq:boundError}. These are the bounds illustrated in \RefFigHighfi{}(e).

\end{document}